\documentclass[times,twocolumn,final,authoryear]{elsarticle}

\usepackage{jasr}
\usepackage{framed,multirow}
\usepackage{float}

\usepackage{amssymb}
\usepackage{latexsym}
\usepackage{nomencl}
\makenomenclature

\usepackage[switch]{lineno}

\usepackage{url}
\usepackage{xcolor}
\definecolor{newcolor}{rgb}{.8,.349,.1}

\journal{}

\begin{document}

\verso{Given-name Surname \textit{etal}}

\begin{frontmatter}

\title{Simplified Optimization Model for Low-Thrust Perturbed Rendezvous Between Low-Eccentricity Orbits}%
\tnotetext[tnote1]{Engineer, Xi'an Satellite Control Center. hay04@foxmail.com}

\author[1]{An-yi \snm{Huang}\corref{cor1}}
\cortext[cor1]{Corresponding author: 
  Tel.: +8684762399;  
  fax: +8684762399;}
\author[1]{Heng-nian \snm{Li}}
\fntext[fn1]{Professor, Xi'an Satellite Control Center.}

\address[1]{State Key Laboratory of Astronautic Dynamics, Xi'an 710043, China}

\received{ }
\finalform{ }
\accepted{ }
\availableonline{ }
\communicated{ }

\begin{abstract}
Trajectory optimization of low-thrust perturbed orbit rendezvous is a crucial technology for space missions in low Earth orbits, which is difficult to solve due to its initial value sensitivity, especially when the transfer trajectory has many revolutions. This paper investigated the time-fixed perturbed orbit rendezvous between low-eccentricity orbits and proposed a priori quasi-optimal thrust strategy to simplify the problem into a parametric optimization problem, which significantly reduces the complexity. The optimal trajectory is divided into three stages including transfer to a certain intermediate orbit, thrust-off drifting and transfer from intermediate orbit to the target orbit. In the two transfer stages, the spacecraft is assumed to use a parametric law of thrust. Then, the optimization model can be then obtained using very few unknowns. Finally, a differential evolution algorithm is adopted to solve the simplified optimization model and an analytical correction process is proposed to eliminate the numerical errors. Simulation results and comparisons with previous methods proved this new method's efficiency and high precision for low-eccentricity orbits. The method can be well applied to premilitary analysis and high-precision trajectory optimization of missions such as in-orbit service and active debris removal in low Earth orbits. 

\end{abstract}


\end{frontmatter}



\section{Introduction}
Trajectory optimization of perturbed orbit rendezvous is a crucial technology for space missions in low Earth orbits (LEOs) \citep{1,2}. Low-thrust electrical propulsion is usually preferred in such missions because of its high efficiency \citep{3,4,5}. As the thrust is insignificant compared with the Earth's gravity, the transfer trajectory usually has many revolutions, which brings additional difficulty for trajectory optimization. In this condition, existing numerical methods \citep{6,7,8,71,72} are very sensitive to the initial values in the shooting process and thus easily converge to local solutions of different revolutions. Moreover, the numerical orbit propagation including the nonlinear perturbations is also time-consuming when the transfer duration is very long. Averaging methods \citep{10,11,12,26} can be applied to replace the time-consuming orbit calculation. However, special assumptions are reqiured and it's hard to find a general optimization method applicable for all type of orbits.\\
This paper mainly studies the fuel-optimal trajectory optimization of time-fixed orbit rendezvous in LEOs with low eccentricity. The major effect of the perturbations on a spacecraft is from the $J_2$ term of the Earth's non-sphere perturbation, which drifts the rising node right ascension (RAAN) and the argument of perigee with constant velocities \citep{9}. Therefore, instead of correcing the derivations brought by $J_2$ perturbation, one can utilize the natural drift of orbit elements actively to save the propellant.\\
Several studies simplified the problem to quickly evaluate the approximate propellant consumption for mission analysis and the global optimization of multi-target rendezvous problems. \cite{13} analyzed the feasibility of actively changing the semimajor axis and inclination to use the natural drift of the right ascension of ascending node and reduce the propellent and studied its application in target selection for an active debris removal mission. \cite{14} discretized the semimajor axis and inclination to search for an optimal RAAN drift rate, which can partly reduce the complexity of global optimization. \cite{15} established an approximate model of low-thrust orbit rendezvous for circular orbits and derived the analytical expression of velocity increment. \cite{16,17} proposed an equality constraint optimization model of impulsive rendezvous and designed an iterative method to expand the impulsive solution to an equivalent low-thrust solution with high precision for orbits with small eccentricity. However, these methods cannot obtain the law of thrust and transfer trajectory.\\
To obtain the approximate thrust law, \cite{18} proposed an optimization model that only considered semimajor axis, inclination, and RAAN based on the minimum principle. \cite{19} improved Cerf's method by introducing the yaw switch strategy and reduced the propellant consumption in some cases. Such idea could be also found in \citep{27,28}. These methods introduced an intermediate drift orbit and let the spacecraft transfer to the drift orbit using Edelbaum's time-optimal strategy \citep{20}. As an improvement, \cite{21} designed a parametric thrust strategy that allows the thrust to periodically switch between on and off when transferring to the drift orbit. Then, an equality constraint optimization model can be obtained and quickly solved. Moreover, in \citep{21}, an analytical correction process was introduced to obtain the high-precision trajectory that considered full perturbations. These methods can only adapt to circular orbits. However, most of the debris and satellites in LEO are in elliptical orbits of small eccentricities, which should be considered in trajectory optimization. Therefore, this study investigates the fast optimization model of low-thrust rendezvous between elliptical orbits. \\
We propose a novel simplified parametric thrust strategy to approximate the optimal control law for fuel-optimal low-thrust rendezvous between low-eccentricity orbits, significantly improving the efficiency of the trajectory optimization. The major contribution can be summarized as three points: \\
(1) Based on the three-stages near-optimal strategy for rendezvous with circular orbits \citep{21}, an approximate optimization model incuding the radial component of thrust and allowing the length of two thrust-on arcs of each revolution to be asymmetric is proposed in this study. Thus, the obtained trajectory could satisfy the constraints on the six-dimentional orbit elements. \\
(2) A fast solving process to judge the feasibility of a low-thrust single-revolution transfer and obtain the thrust parameters is proposed to reduce the dimensionality of the optimization model and improve the efficiency.  \\
(3) A fast analytical correction process is proposed to obtain high-precision trajectory using the numerical errors between predicted orbit and target orbit.\\
Simulation results proved that the proposed method could quickly obtain the optimal high-precision transfer trajectory. The calculation is much smaller than indirect methods that often obtain local optimal solutions and thus need to repeat the shooting process with different initial values of costate for selecting the best solution. Compared with existing approximate methods, the proposed parametric optimization method considers constraints on eccentricity and is more precise for elliptical orbits.\\

\section{Problem description}
This study focuses on the fuel-optimal trajectory of time-fixed orbital rendezvous. Assuming ${\sigma _0} = [{a_0},{e_0},{i_0},{\Omega _0},{\omega _0},{M_0}]$ represents the initial orbit elements (a: semimajor axis, e: eccentricity, i: inclination, $\Omega$: RAAN, $\omega$: argument of perigee, M: mean anomaly, and subscript ‘0’ means initial orbit) for the spacecraft at ${t_0}$ and ${\sigma _f} = [{a_f},{e_f},{i_f},{\Omega _f},{\omega _f},{M_f}]$ (subscript ‘f’ means target orbit) presents the target orbit elements at ${t_f}$, the trajectory optimization is a typical optimal control problem that need to solve the optimal thrust acceleration ${\bf{\alpha }}(t)$ that satisfy the terminal constraint and minimize the propellent cost. \\
Let $\sigma (t) = [a,e,i,\Omega ,\omega ,M]$ denote the orbit elements during the transfer, the dynamics equations are
\begin{footnotesize}
\begin{equation}
\left\{\begin{array}{l}
\frac{{{\rm{d}}a}}{{{\rm{d}}t}} = \frac{2}{{n\sqrt {1 - {e^2}} }}[({\alpha _r} + \alpha _r^{{J_2}})e\sin f + ({\alpha _t} + \alpha _t^{{J_2}})(1 + e\cos f)]\\
\frac{{{\rm{d}}e}}{{{\rm{d}}t}} = \frac{{\sqrt {1 - {e^2}} }}{{na}}[({\alpha _r} + \alpha _r^{{J_2}})\sin f + ({\alpha _t} + \alpha _t^{{J_2}})(\cos E + \cos f)]\\
\frac{{{\rm{d}}i}}{{{\rm{d}}t}} = \frac{{r\cos u}}{{na\sqrt {1 - {e^2}} }}({\alpha _n} + \alpha _n^{{J_2}})\\
\frac{{{\rm{d}}\Omega }}{{{\rm{d}}t}} = \frac{{r\sin u}}{{na\sqrt {1 - {e^2}} \sin i}}({\alpha _n} + \alpha _n^{{J_2}})\\
\frac{{{\rm{d}}\omega }}{{{\rm{d}}t}} = \frac{{\sqrt {1 - {e^2}} }}{{nae}}[ - ({\alpha _r} + \alpha _r^{{J_2}})\cos f + ({\alpha _t} + \alpha _t^{{J_2}})(1 + \frac{r}{p})\sin f] - \cos i\frac{{{\rm{d}}\Omega }}{{{\rm{d}}t}}\\
\frac{{{\rm{d}}M}}{{{\rm{d}}t}} = n - \frac{{1 - {e^2}}}{{nae}}[({\alpha _r} + \alpha _r^{{J_2}})(2e\frac{r}{p} - \cos f) + ({\alpha _t} + \alpha _t^{{J_2}})(1 + \frac{r}{p})\sin f]
\end{array}\right.
\end{equation}
\end{footnotesize}
where $f$ is the true anomaly, $p = a(1 - {e^2})$ is the semi-latus rectum, $u$ is the argument of latitude, $n = \sqrt {\frac{\mu }{{{a^3}}}} $ is the orbital angular velocity, and ${\bf{\alpha }}_{}^{{J_2}} = [\alpha _t^{{J_2}},\alpha _n^{{J_2}},\alpha _r^{{J_2}}]$ are the three acceleration components due to the J2 perturbation in the local vertical/local horizontal (LVLH) reference frame. ${\alpha _t} = c(t)\alpha \cos \beta ,{\alpha _n} = c(t)\alpha \sin \beta \cos \phi $and${\alpha _r} = c(t)\alpha \sin \beta \sin \phi $ are the three components of ${\bf{\alpha }}(t)$ in the LVLH reference frame, where $\beta $ is the angle between ${\bf{\alpha }}(t)$ and the tangential direction, $\phi $ is the angle between the projection of ${\bf{\alpha }}(t)$ in the normal-radial plane and the radial direction, $\alpha  = \frac{{{F_{\max }}}}{m}$ is the max acceleration expressed by the maximum thrust ${F_{\max }}$ and mass $m$, and $c(t) \in [0,1]$ is engine throttling function representing magnitude of the thrust acceleration.\\
The objective function can be written as
\begin{equation}
J = \dot m\int_{{t_0}}^{{t_f}} {c(t)dt} 
\end{equation}
where $\dot m = \frac{F}{{{I_{sp}}g}}$ is the constant mass flow rate, ${I_{sp}}$ is the constant specific impulse, and $g$ is the standard gravitational acceleration at sea level (9.80665 m/s2).\\
In this study’s investigation, when the propellant cost is small compared to the mass (for example, when $I_{sp}$ = 1000 s and $\Delta v$ = 200 m/s, the fuel cost is apprxoimately 2\%), it can be supposed that the mass is constant during the orbit transfer. Then, we can let $\alpha  = \frac{{{F_{\max }}}}{{{m_0}}}$ denote the maximum acceleration (${m_0}$ is the initial mass). Since $\alpha $ is much smaller than the Earth's gravity, completing the orbit rendezvous requires a long time. The number of orbit revolutions would be significant, and the orbit propagation would be time-consuming. In the next section, a simplified thrust strategy will be proposed to express $c(t)$, $\beta (t)$, and $\phi (t)$ by a few parameters to reduce the complexity. \\

\section{Methodology}
This section first designs a parametric thrust strategy for orbit rendezvous with elliptical orbit and then establishes the simplified optimization model. A sub boundary value problem that solves the optimal parameters corresponding to orbital element changes in a single revolution is embedded in the optimization model to resolve terminal constraints. Finally, a differential evolution algorithm is employed to solve the optimal trajectory.\\
\subsection{Near-optimal thrust strategy}
\begin{figure}[hbt!]
\centering
\includegraphics[scale=1.4]{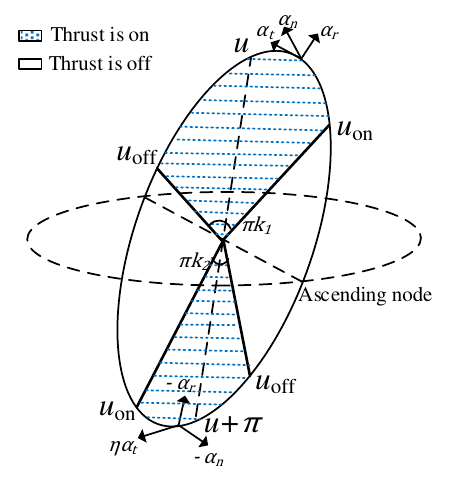}
\caption{Fixed thrust strategy in one revolution.}
\label{fig:fig1}
\end{figure}

A near-optimal thrust strategy \citep{21} is proposed for transfer between circular orbits and has shown to be efficient. It divides the transfer duration into three stages: transfer to arrive at an intermediate orbit with certain RAAN drift, natural drift duration with no thrust, and transfer from intermediate orbit to the target orbit. Thus, most of the non-coplanar maneuvers required for RAAN control can be avoided with the help of natural RAAN drift caused by J2 perturbation. In the first and third stages, it’s assumed that the near-optimal Bang-Bang control law switches the thrust on and off periodically in each revolution. When the thrust is on, its direction is assumed to be fixed in the LVLH reference frame. Then, the low-thrust optimization problem is simplified using very few parameters \citep{21}.\\
In this study, to satisfy the eccentricity constraints for orbit rendezvous, we expand the thrust strategy by including the radial thrust component to change the eccentricity jointly with the tangential component. As illustrated in Fig. \ref{fig:fig1}, there are two thrust arcs in each revolution, and their middle points are symmetric. When the thrust is on, the acceleration keeps the maximum value. The lengths of the two arcs could not be equal to change eccentricity by the tangential component of thrust while keeping the semimajor axis unchanged. Let ${u}$ and ${u} + \pi $ denote the arguments of altitude, and $\pi {k_1}$ and $\pi {k_2}$ denote the lengths of the two arcs. The three thrust components are fixed ($\beta $ and $\phi $ are constant) in the LVLH reference frame, and the values of radial and normal components are opposite in the two arcs. The sign of the tangential components may be the same or opposite, which is expressed by the unknown coefficient $\eta  =  - 1 or 1$. Then, the trajectory of one revolution can be determined using the six parameters $\eta $,${k_1}$,${k_2}$,$u$,$\beta $, and $\phi $.\\ 
Then, based on the three-stage near-optimal thrust strategy \citep{21} and the improvment above for eccentricity control, the whole transfer trajectory can be expressed by 14 parameters: $\Delta {t_1}$ and $\Delta {t_2}$ denote the duration of the first and third stages; ${\eta _1}$,${k_{11}}$,${k_{12}}$,${u_1}$,${\beta _1}$, and ${\phi _1}$ denote the parameters of thrust in the first stage (using additional subscript '1'); and ${\eta _2}$,${k_{21}}$,${k_{22}}$,${u_2}$,${\beta _2}$, and ${\phi _2}$ denote the parameters of thrust in the third stage (using additional subscript '2'). The parametric orbit rendezvous problem is detailed in the following subsections.\\
\subsection{Assumptions declaration}
To obtain a simplified model, several assumptions should be declared. \\
(1) First, the initial and target orbits are near-circular orbits, and $p \approx a \approx r$. In addition, the changes in the semimajor axis and inclination are small enough, thus the $a$, $i$, and $n$ in the right function of Eq. (1) are approximately constant and equal to their initial values ${a_0}$, ${i_0}$ and ${n_0}$. \\
Here, ${\rm{d}}t$ in Eq. (1) can be replaced by ${\rm{d}}u = {n_0}{\rm{d}}t$ because $n$ is also near constant. e and $\omega $ are replaced by ${e_x} = e\cos \omega $ and ${e_y} = e\sin \omega $ to avoid a singularity. Then, Eq. (1) can be divided into two independent parts: effects of thrust and $J_2$ perturbation. The changes in orbit elements by the thrust (without perturbation) are:
\begin{equation}
\left\{\begin{array}{l}
\frac{{{\rm{d}}a}}{{{\rm{d}}u}} = \frac{{2{a_0}\alpha \cos \beta }}{{{n_0}{V_0}}}\\
\frac{{{\rm{d}}i}}{{{\rm{d}}u}} = \frac{{\alpha \sin \beta \cos \phi \cos u}}{{{n_0}{V_0}}}\\
\frac{{{\rm{d}}\Omega }}{{{\rm{d}}u}} = \frac{{\alpha \sin \beta \cos \phi \sin u}}{{{n_0}{V_0}\sin i}}\\
\frac{{{\rm{d}}{e_x}}}{{{\rm{d}}u}} = \frac{\alpha }{{{n_0}{V_0}}}(2\cos \beta \cos u - \sin \beta \sin \phi \sin u)\\
\frac{{{\rm{d}}{e_y}}}{{{\rm{d}}u}} = \frac{\alpha }{{{n_0}{V_0}}}(2\cos \beta \sin u - \sin \beta \sin \phi \cos u)
\end{array}\right.
\end{equation}
where ${V_0} = {n_0}{a_0}$ is the mean orbital velocity.\\
The effect of $J_2$ perturbation is expressed by the analytical form in \citep{25} as follows:
\begin{equation}
\left\{\begin{array}{l}
\frac{{{\rm{d}}a}}{{{\rm{d}}t}} = 0\\
\frac{{{\rm{d}}i}}{{{\rm{d}}t}} = 0\\
\frac{{{\rm{d}}\Omega }}{{{\rm{d}}t}} =  - \frac{{3{J_2}{n_0}r_E^2\cos {i_0}}}{{2{a_0}^2}}\\
\frac{{{\rm{d}}{e_x}}}{{{\rm{d}}t}} =  - e\sin \omega \frac{{{\rm{d}}\omega }}{{{\rm{d}}t}}\\
\frac{{{\rm{d}}{e_y}}}{{{\rm{d}}t}} = e\cos \omega \frac{{{\rm{d}}\omega }}{{{\rm{d}}t}}\\
\frac{{{\rm{d}}\omega }}{{{\rm{d}}t}} =  - \frac{{3{J_2}{n_0}r_E^2(2 - 2.5{{\sin }^2}{i_0})}}{{2{a_0}^2}}
\end{array}\right.
\end{equation}
where $r_E^{}$ is the mean equator radius.\\
Note that according to Eq. (2) and Eq. (3), when calculating the effects of thrust and perturbations, we assumed the orbit is circular. The bias of this assumption can be evaluated by the terms in Eq. (1) that include $e$ and $p$. Assume $e = 0.1$ (for most near-circular satellites in LEO), then, $p = 0.99 a$ and the gap can be ignored. Meanwhile, the maximum error of $e\cos f$ and $e\sin f$ is 10\% and the accumulate error would be much less after an integral from 0 to $2\pi$. The impulsive trajectory optimization in \citep{16} used the same assumption and the simulation indicated that although the eccentricity difference is a little greater than 0.1, the relative error is less than 5\%. Moreover, we will also provide iteration process in the next section to correct the law of thrust and eliminate the deviations of terminal states. Therefore, the assumption would be reasonable and applicable for most cases in LEO. \\
(2) Second, as illustrated in Fig.\ref{fig:fig1}, $c(u)$, ${\alpha _n}$, ${\alpha _t}$, and ${\alpha _r}$ are defined as in Eqs. (5) and (6), when ${\eta _{}}$,${k_1}$,${k_2}$,${u_{}}$,${\beta _{}}$, and ${\phi _{}}$ are given. \\
\begin{equation}
c(u) = \left\{ \begin{array}{l}
1,{\rm{ if }}u \in [ - \frac{{{k_1}\pi }}{2} + {u_1},\frac{{{k_1}\pi }}{2} + {u_1}]\\
1,{\rm{ if }}u \in [ - \frac{{{k_2}\pi }}{2} + {u_1} + \pi ,\frac{{{k_2}\pi }}{2} + {u_1} + \pi ]\\
0,{\rm{ else}}
\end{array} \right.
\end{equation}
\begin{footnotesize}
\begin{equation}
\begin{array}{l}
{\alpha _t} = \left\{ \begin{array}{l}
\alpha \cos \beta ,{\rm{if }}u \in [ - \frac{{{k_1}\pi }}{2} + {u_1},\frac{{{k_1}\pi }}{2} + {u_1}]\\
\eta \alpha \cos \beta ,{\rm{if }}u \in [ - \frac{{{k_2}\pi }}{2} + {u_1} + \pi ,\frac{{{k_2}\pi }}{2} + {u_1} + \pi ]
\end{array} \right.\\
{\alpha _n} = \left\{ \begin{array}{l}
\alpha \sin \beta \cos \phi ,{\rm{if }}u \in [ - \frac{{{k_1}\pi }}{2} + {u_1},\frac{{{k_1}\pi }}{2} + {u_1}]\\
 - \alpha \sin \beta \cos \phi ,{\rm{if }}u \in [ - \frac{{{k_2}\pi }}{2} + {u_1} + \pi ,\frac{{{k_2}\pi }}{2} + {u_1} + \pi ]
\end{array} \right.\\
{\alpha _r} = \left\{ \begin{array}{l}
\alpha \sin \beta \sin \phi ,{\rm{if }}u \in [ - \frac{{{k_1}\pi }}{2} + {u_1},\frac{{{k_1}\pi }}{2} + {u_1}]\\
 - \alpha \sin \beta \sin \phi ,{\rm{if }}u \in [ - \frac{{{k_2}\pi }}{2} + {u_1} + \pi ,\frac{{{k_2}\pi }}{2} + {u_1} + \pi ]
\end{array} \right.
\end{array}
\end{equation}
\end{footnotesize}
where ${k_1}$ and ${k_2}$ should be positive and ${k_1}\pi  + {k_2}\pi  \leq 2\pi $.\\
Then, according to Eq. (3), the changes in elements after one revolution is the definite integral from 0 to $2\pi $, which is an extension of Eq. (3) in \citep{21} by involving the eccentricity: 
\begin{footnotesize}
\begin{equation}
\left\{\begin{array}{l}
\Delta a = \int_{{u_1} - {k_1}\pi /2}^{{u_1} + {k_1}\pi /2} {{\rm{d}}a}  + \int_{{u_1} + \pi  - {k_2}\pi /2}^{{u_1} + \pi  + {k_2}\pi /2} {{\rm{d}}a}  = \frac{{{a_0}\alpha \cos \beta }}{{{V_0}}}({k_1} + \eta {k_2})T\\
\Delta i = \int_{{u_1} - {k_1}\pi /2}^{{u_1} + {k_1}\pi /2} {{\rm{d}}i}  + \int_{{u_1} + \pi  - {k_2}\pi /2}^{{u_1} + \pi  + {k_2}\pi /2} {{\rm{d}}i}  = \frac{{({k_1}k_1^T + {k_2}k_2^T)\alpha \sin \beta \cos \phi \cos {u_1}}}{{2{V_0}}}T\\
\Delta \Omega  = \int_{{u_1} - {k_1}\pi /2}^{{u_1} + {k_1}\pi /2} {{\rm{d}}\Omega }  + \int_{{u_1} + \pi  - {k_2}\pi /2}^{{u_1} + \pi  + {k_2}\pi /2} {{\rm{d}}\Omega }  = \frac{{({k_1}k_1^T + {k_2}k_2^T)\alpha \sin \beta \cos \phi \cos {u_1}}}{{2{V_0}\sin i}}T\\
\Delta {e_x} = \int_{{u_1} - {k_{11}}\pi /2}^{{u_1} + {k_{11}}\pi /2} {{\rm{d}}{e_x}}  + \int_{{u_1} + \pi  - {k_2}\pi /2}^{{u_1} + \pi  + {k_2}\pi /2} {{\rm{d}}{e_x}} \\
 = [\frac{{({k_1}k_1^T - \eta {k_2}k_2^T)\alpha \cos \beta \cos {u_1}}}{{{V_0}}} + \frac{{({k_1}k_1^T + {k_2}k_2^T)\alpha \sin \beta \sin \phi \sin {u_1}}}{{2{V_0}}}]T\\
\Delta {e_y} = \int_{{u_1} - {k_{11}}\pi /2}^{{u_1} + {k_{11}}\pi /2} {{\rm{d}}{e_y}}  + \int_{{u_1} + \pi  - {k_{12}}\pi /2}^{{u_1} + \pi  + {k_{12}}\pi /2} {{\rm{d}}{e_y}} \\
 = [\frac{{({k_1}k_1^T - \eta {k_2}k_2^T)\alpha \cos \beta \sin {u_1}}}{{{V_0}}} - \frac{{({k_{11}}k_{11}^T + {k_{12}}k_{12}^T)\alpha \sin \beta \sin \phi \cos {u_1}}}{{2{V_0}}}]T
\end{array} \right.
\end{equation}
\end{footnotesize}
where $\alpha  = {\alpha _{\max }}$,$T = \frac{{2\pi }}{n}$ is the orbit period, $k_1^T = \frac{{\sin \frac{{\pi {k_1}}}{2}}}{{\pi {k_1}/2}}$, and $k_2^T = \frac{{\sin \frac{{\pi {k_2}}}{2}}}{{\pi {k_2}/2}}$ are the equivalent coefficients of thrust corresponding to ${k_1}$ and ${k_2}$.\\
When the transfer duration is much longer than T, assuming that the orbital elements change at a uniformly average speed is reasonable \citep{21, 22}. Thus, orbit changes after a given time $\Delta t$ can be calculated by replacing T with $\Delta t$ in Eq. (7). \\
(3) Third, according to Eq. (4) the changes in the semimajor axis and inclination will lead to accumulative changes in the argument of latitude, the argument of perigee, and RAAN, which can be analytically calculated as follows.\\
\begin{figure}[hbt!]
\centering
\includegraphics[scale=0.7]{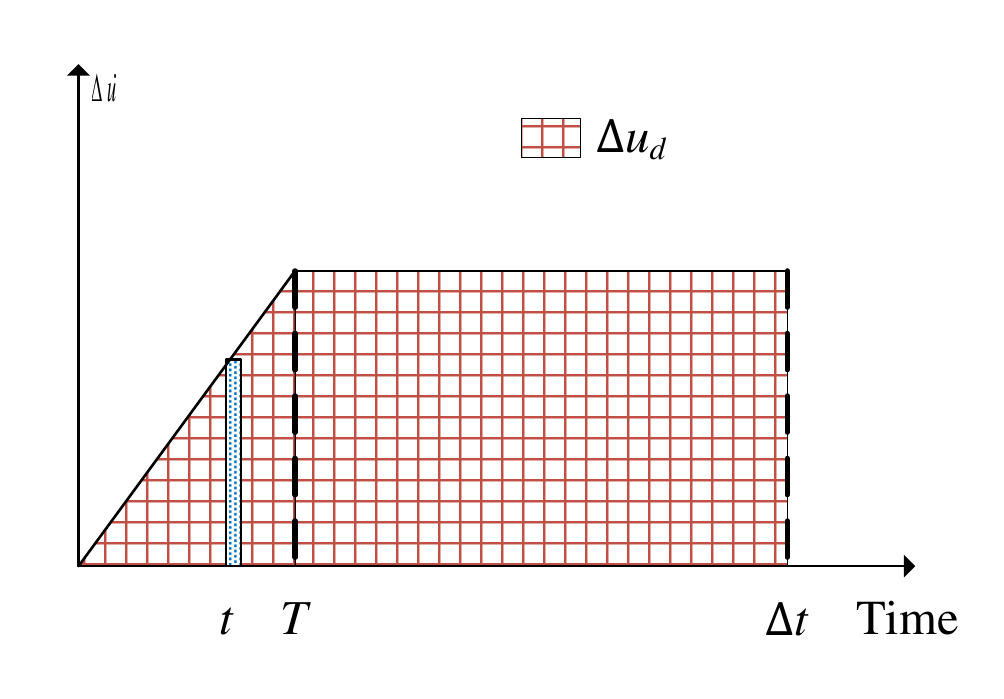}
\caption{Drift of argument of latitude.}
\label{fig:fig2}
\end{figure}
Assuming that $\Delta a$ and $\Delta i$ are obtained using Eq. (7) and the orbital period is $T$. Then, the average change rate of $a$ and $i$ are $\dot a = \frac{{\Delta a}}{T}$ and $\dot i = \frac{{\Delta i}}{T}$. Although we assume $u$ moves at a constant velocity $n_0$ when calculating other orbit elements by Eq. (7), the effect of $\Delta a$ on $u$ after a long duration must be considered for accurate orbit rendezvous, as illustarted in Fig. \ref{fig:fig2}($\Delta \dot u$ is the relative drfit rate of $u$ compared with the initial orbit). The change in $u$ after a given duration $\Delta t$ (including $T$) can be expressed by the definite integral as:
\begin{equation}
\begin{array}{l}
\Delta {u_d} = \int_0^{\Delta t} {\Delta \dot u{\rm{d}}t = } \int_0^T {\frac{{{\rm{d}}\dot u}}{{{\rm{d}}a}}\dot at{\rm{d}}t}  + (\Delta t - T)\frac{{{\rm{d}}\dot u}}{{{\rm{d}}a}}\dot aT\\
 =  - \frac{{3n\dot a}}{{2a}}(\frac{1}{2}{T^2} + (\Delta t - T)T)\\
 =  - \frac{{3n\Delta a}}{{2a}}(\Delta t - \frac{1}{2}T)\\
 = ({n_T} - {n_0})(\Delta t - \frac{1}{2}T)
\end{array}
\end{equation}
where we assume $\Delta \dot u \approx \frac{{{\rm{d}}\dot u}}{{{\rm{d}}a}}\Delta a$. $\frac{{{\rm{d}}\dot u}}{{{\rm{d}}a}} = \frac{{{\rm{d}}n}}{{{\rm{d}}a}} =  - \frac{{3n}}{{2a}}$ represents the derivative of $u$ with respect to semimajor axis. $n_0$ and $n_T$ are the angular velocities of initial orbit and orbit after one-revolution control.\\
Similarly, the change in RAAN after $\Delta t$ is
\begin{equation}
\begin{array}{*{20}{l}}
{\Delta {\Omega _d} = \int_0^{\Delta t} {\Delta \dot \Omega {\rm{d}}t} }\\
{ = \int_0^T {(\frac{{{\rm{d}}\dot \Omega }}{{{\rm{d}}a}}\dot at + \frac{{{\rm{d}}\dot \Omega }}{{{\rm{d}}i}}\dot it){\rm{d}}t}  + (\Delta t - T)(\frac{{{\rm{d}}\dot \Omega }}{{{\rm{d}}a}}\dot aT + \frac{{{\rm{d}}\dot \Omega }}{{{\rm{d}}i}}\dot iT)}\\
\begin{array}{l}
 = (\frac{{{\rm{d}}\dot \Omega }}{{{\rm{d}}a}}\dot a + \frac{{{\rm{d}}\dot \Omega }}{{{\rm{d}}i}}\dot i)(\frac{{{T^2}}}{2} + \Delta tT - {T^2})\\
 = \Delta {{\dot \Omega }_T}(\Delta t - \frac{T}{2})
\end{array}
\end{array}
\end{equation}
where we assume $\Delta \dot \Omega  \approx \frac{{{\rm{d}}\dot \Omega }}{{{\rm{d}}a}}\Delta a + \frac{{{\rm{d}}\dot \Omega }}{{{\rm{d}}i}}\Delta i$. $\frac{{{\rm{d}}\dot \Omega}}{{{\rm{d}}a}}$ and $\frac{{{\rm{d}}\dot \Omega}}{{{\rm{d}}i}}$ are the derivative of $\dot \Omega$ with respect to semimajor axis and inclination by Eq.(4). ${\Delta \dot \Omega _T} = {\dot \Omega _T} - {\dot \Omega _0}$ represents the relative drift rate of RAAN. ${\dot \Omega _0}$ and ${\dot \Omega _T}$ are the RAAN drift rates of initial orbit and the orbit after $T$, respectively.\\
The change in $\omega $ after $\Delta t$ is
\begin{equation}
\begin{array}{l}
\Delta {\omega _d} = \int_0^{\Delta t} {\Delta \dot \omega {\rm{d}}t} \\
 = \int_0^T {(\frac{{{\rm{d}}\dot \omega }}{{{\rm{d}}a}}\dot at + \frac{{{\rm{d}}\dot \omega }}{{{\rm{d}}i}}\dot it){\rm{d}}t}  + (\Delta t - T)(\frac{{{\rm{d}}\dot \omega }}{{{\rm{d}}a}}\dot aT + \frac{{{\rm{d}}\dot \omega }}{{{\rm{d}}i}}\dot iT)\\
 = {{\Delta \dot \omega }_T}(\Delta t - \frac{T}{2})
\end{array}
\end{equation}
where we assume $\Delta \dot \omega  \approx \frac{{{\rm{d}}\dot \omega }}{{{\rm{d}}a}}\Delta a + \frac{{{\rm{d}}\dot \omega }}{{{\rm{d}}i}}\Delta i$. $\frac{{{\rm{d}}\dot \omega}}{{{\rm{d}}a}}$ and $\frac{{{\rm{d}}\dot \omega}}{{{\rm{d}}i}}$ are the derivative of $\dot \omega$ with respect to semimajor axis and inclination by Eq.(4). ${\Delta \dot \omega _T} = {\dot \omega _T} - {\dot \omega _0}$ represents the relative drift rate of argument of perigee. ${\dot \omega _0}$ and ${\dot \omega _T}$ are the drift rates of initial orbit and the orbit after $T$, respectively.\\
(4) Four, the propellant cost is sufficiently small compared to the spacecraft's mass, thus the fuel-optimal objective function is equal to minimizing the actual time of thrust-on arcs. The length of thrust-on arcs in one revolution is calculated as
\begin{equation}
\Delta {t_{thrust}} = \frac{{({k_1} + {k_2})}}{2}T
\end{equation}

\subsection{Optimization model}
According to Eqs. (3) ~ (11), the terminal constraints $\sigma ({t_f}) = {\sigma _f}$ and objective function can be analytically expressed by the 14-dimensional unkowns ($\Delta {t_1},\Delta {t_2},{\eta _1},{k_1}_1,{k_1}_2,{\beta _1},{\phi _1},{u_1}$,${\eta _2},{k_2}_1,{k_2}_2,{\beta _2},{\phi _2},$ and ${u_2}$). 
First, the changes in orbit elements during the first stage $\Delta {t_1}$ (transfer from the initial orbit to the intermediate drift orbit) are
\begin{footnotesize}{small}
\begin{equation}
\left\{\begin{array}{l}
\Delta {a_1} = \frac{{\alpha \cos \beta {a_0}}}{{{V_0}}}({k_{11}} + \eta {k_{12}})\Delta {t_1}\\
\Delta {i_1} = \frac{{({k_{11}}k_{11}^T + {k_{12}}k_{12}^T)\alpha \sin {\beta _1}\cos {\phi _1}\cos {u_1}}}{{2{V_0}}}\Delta {t_1}\\
\Delta {\Omega _1} = \frac{{({k_{11}}k_{11}^T + {k_{12}}k_{12}^T)\alpha \sin {\beta _1}\cos {\phi _1}\sin {u_1}}}{{2{V_0}\sin {i_0}}}\Delta {t_1}\\
\Delta {e_x}_1 = [\frac{{({k_{11}}k_{11}^T - \eta {k_{12}}k_{12}^T)\alpha \cos {\beta _1}\cos {u_1}}}{{{V_0}}} + \frac{{({k_{11}}k_{11}^T + {k_{12}}k_{12}^T)\alpha \sin {\beta _1}\sin {\phi _1}\sin {u_1}}}{{2{V_0}}}]\Delta {t_1}\\
\Delta {e_y}_1 = [\frac{{({k_{11}}k_{11}^T - \eta {k_{12}}k_{12}^T)\alpha \cos {\beta _1}\sin {u_1}}}{{{V_0}}} - \frac{{({k_{11}}k_{11}^T + {k_{12}}k_{12}^T)\alpha \sin {\beta _1}\sin {\phi _1}\cos {u_1}}}{{2{V_0}}}]\Delta {t_1}
\end{array}\right.
\end{equation}
\end{footnotesize}
Similarly, the changes in orbit elements $\Delta {a_2},\Delta {i_2},\Delta {\Omega _2},\Delta {e_x}_2,\Delta {e_y}_2$ during $\Delta {t_2}$(transfer from the intermediate drift orbit to the target orbit) can be calculated. To complete the orbit rendezvous, the constraints are:
\begin{equation}
\left\{\begin{array}{l}
\Delta {a_2} + \Delta {a_1} = \Delta {a_0}\\
\Delta {i_2} + \Delta {i_1} = \Delta {i_0}\\
\Delta {\Omega _2} + \Delta {\Omega _1} + \Delta {\Omega _d} = \Delta {\Omega _0}\\
\Delta {e_x}_2 + \Delta {e_x}_1' = \Delta {e_x}_0\\
\Delta {e_y}_2 + \Delta {e_{y1}}' = \Delta {e_y}_0\\
\Delta {u_d} = \Delta {u_0}
\end{array}\right.
\end{equation}
\begin{figure}[hbt!]
\centering
\includegraphics[scale=0.7]{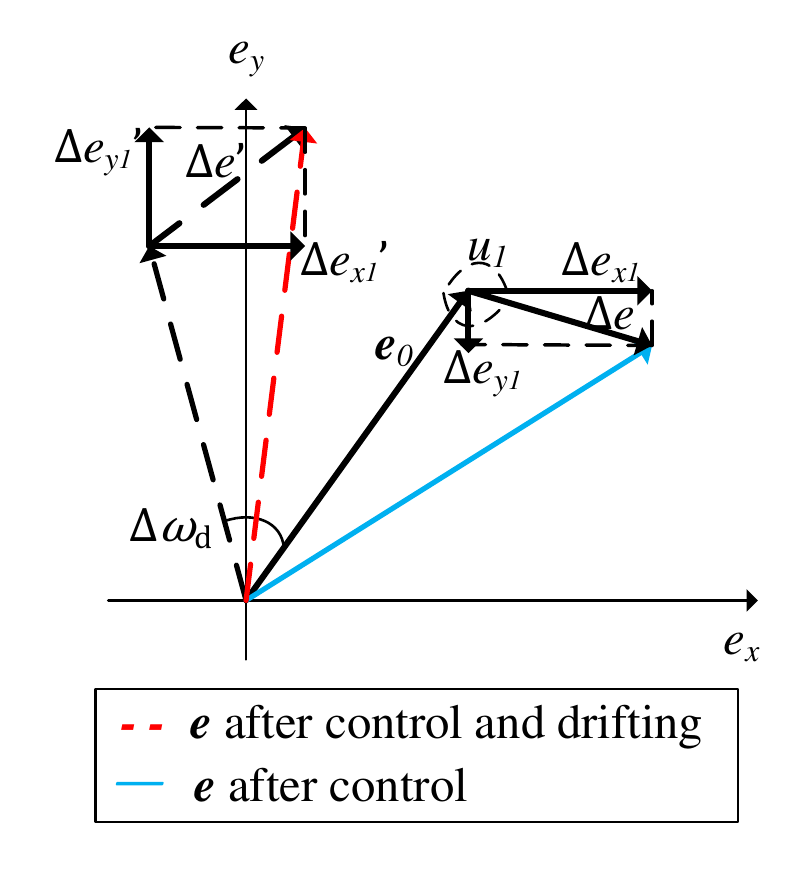}
\caption{Drift of eccentricity change.}
\label{fig:fig3}
\end{figure}
\begin{figure}[hbt!]
\centering
\includegraphics[scale=0.7]{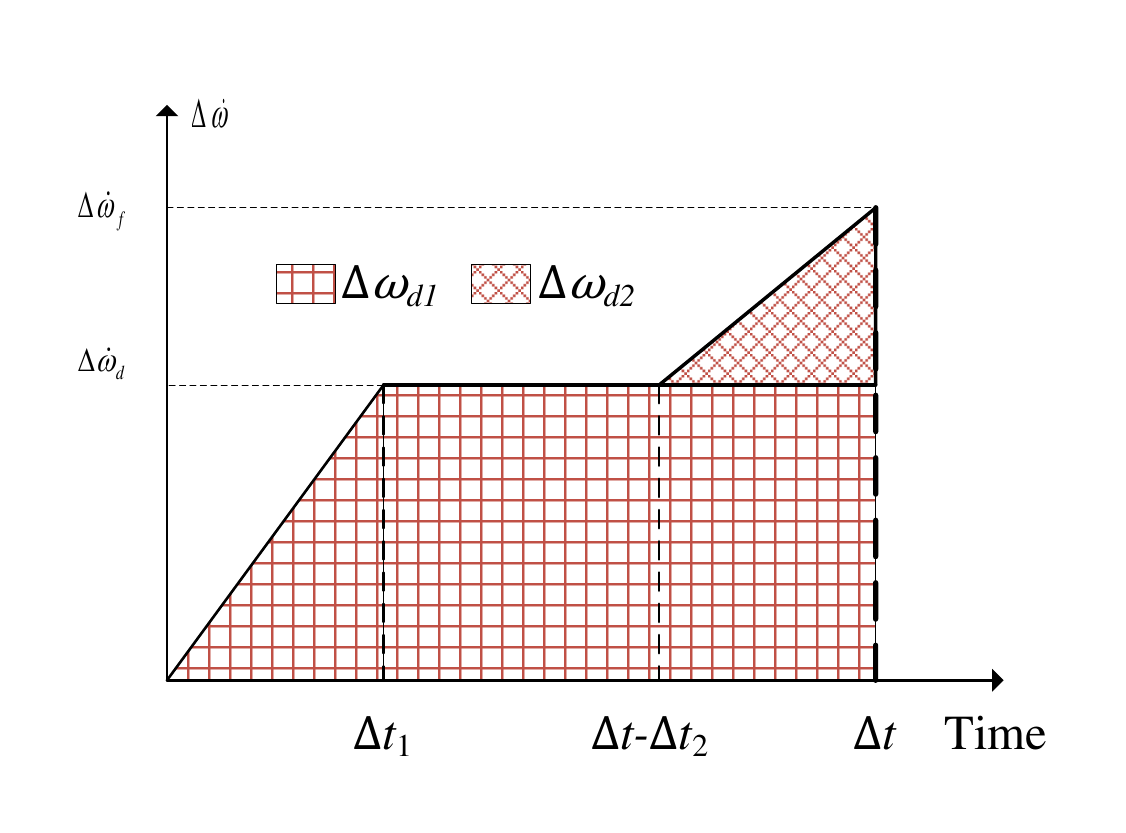}
\caption{Calculation of $\Delta \omega_d$.}
\label{fig:fig4}
\end{figure}
where $\Delta {a_0},\Delta {i_0},\Delta {\Omega _0},\Delta {u_0},\Delta {e_{x0}},\Delta {e_{y0}}$ represent the orbit differences between the initial and target orbit. $\Delta {e_x}_1'$ and $\Delta {e_{y1}}'$ are the corrected changes in eccentricity when considering the drifting of argument of perigee after $\Delta t$ (as illustrated in Fig. \ref{fig:fig3}):\\
\begin{equation}
\left\{\begin{array}{l}
\Delta {e_x}_1' = [\frac{{({k_{11}}k_{11}^T - \eta {k_{12}}k_{12}^T)\alpha \cos {\beta _1}\cos ({u_1} + \Delta {\omega _d})}}{{{V_0}}} \\- \frac{{({k_{11}}k_{11}^T + {k_{12}}k_{12}^T)\alpha \sin {\beta _1}\sin {\phi _1}\sin ({u_1} + \Delta {\omega _d})}}{{2{V_0}}}]\Delta {t_1}\\
\Delta {e_{y1}}' = [\frac{{({k_{11}}k_{11}^T - \eta {k_{12}}k_{12}^T)\alpha \cos {\beta _1}\sin ({u_1} + \Delta {\omega _d})}}{{{V_0}}} \\- \frac{{({k_{11}}k_{11}^T + {k_{12}}k_{12}^T)\alpha \sin {\beta _1}\sin {\phi _1}\cos ({u_1} + \Delta {\omega _d})}}{{2{V_0}}}]\Delta {t_1}
\end{array}\right.
\end{equation}
where $\Delta {\omega _d}$ means the rotate angle of $\Delta {e_x}_1$ and $\Delta {e_{y1}}$ from beginning to $\Delta t$ (as illustrated in Fig. \ref{fig:fig4}) and can be calculated by substituting $\Delta {t_1}$ and $\Delta {t_2}$ to Eq. (10) to replace $T$ and summarizing the results:
\begin{equation}
\begin{array}{l}
\Delta {\omega _d} = \Delta {\omega _{d1}} + \Delta {\omega _{d2}}\\
 = ({{\dot \omega }_d} - {{\dot \omega }_0})(\Delta t - \frac{{\Delta {t_1}}}{2}) + ({{\dot \omega }_f} - {{\dot \omega }_d})(\Delta {t_2} - \frac{{\Delta {t_2}}}{2})
\end{array}
\end{equation}
where ${\dot \omega _0}$,${\dot \omega _d}$, and ${\dot \omega _f}$ are the drift rates of $\omega $ of the initial orbit, intermediate orbit, and target orbits. $\Delta {e_{x2}}$ and $\Delta {e_{y2}}$ do not need to be corrected because the third transfer stage is close to the rendezvous time and the drift of $\omega$ can be ignored.\\
In the same way as Fig. \ref{fig:fig3}, $\Delta {\Omega _d}$ in Eq. (13) represents the change in $\Omega $ by the perturbation and can be calculated by Eq. (9):
\begin{equation}
\Delta {\Omega _d} = ({\dot \Omega _d} - {\dot \Omega _0})(\Delta t - \frac{{\Delta {t_1}}}{2}) + ({\dot \Omega _f} - {\dot \Omega _d})(\Delta {t_2} - \frac{{\Delta {t_2}}}{2})
\end{equation}
where ${\dot \Omega _0}$,${\dot \Omega _d}$, and ${\dot \Omega _f}$ are the drift rates of $\Omega $ of the initial orbit, intermediate orbit, and target orbits.\\
In the same way, $\Delta {u_d}$ in Eq. (13) represents the change in the argument of latitude caused by the semimajor axis and can be calculated by Eq. (8):
\begin{equation}
\Delta {u_d} = ({n_d} - {n_0})(\Delta t - \frac{{\Delta {t_1}}}{2}) + ({n_f} - {n_d})(\Delta {t_2} - \frac{{\Delta {t_2}}}{2})
\end{equation}
where ${n_0}$, ${n_d}$, and ${n_f}$ are the angular velocities of the initial orbit, intermediate orbit, and target orbits, respectively. Since ${n_d}$ is only determined by $\Delta {a_1}$ ($n_d$), when ${\eta _1},{u_1},{k_{11}},{k_{12}},{\beta _1},{\phi _1}$ are given, $\Delta {u_d}$ can be directly calculated by Eq. (17) and may be not equal to $\Delta {u_0}$. Then, we can always add a correction term to ${k_{11}}$ by Eq. (18) to ensure $\Delta {u_d} = \Delta {u_0}$:
\begin{equation}
\Delta {k_{11}} =  - \frac{2}{3}\frac{{\Delta {u_0} - \Delta {u_d}}}{{\Delta t{n_0}}}\frac{{{V_0}}}{{\alpha \cos \beta \Delta {t_1}}}
\end{equation}
where $\Delta {k_{11}}$ is the correction to ${k_{11}}$. Thus, the constraint of orbit rendezvous on $u$ is automatically satisfied.\\
The objective function is written as
\begin{equation}
J = \frac{{({k_{11}} + {k_{12}})}}{2}\Delta {t_1} + \frac{{({k_{21}} + {k_{22}})}}{2}\Delta {t_2}
\end{equation}
Above all, Eqs. (13) to (19) form the optimization model of 14 parameters and 5 constraints. According to the possible values of ${\eta _1}$ and ${\eta _2}$, there are 4 conditions (${\eta _1} = 1,{\eta _2} = 1$; ${\eta _1} =  - 1,{\eta _2} = 1$; ${\eta _1} = 1,{\eta _2} =  - 1$; and ${\eta _1} =  - 1,{\eta _2} =  - 1$) that can be solved separately and the solution that minimizes J can be chosen as the optimal solution. \\

\subsection{Dimensionality reduction via a sub boundary value problem}
Note that $\Delta {a_2},\Delta {i_2},\Delta {\Omega _2},\Delta {e_{x2}},\Delta {e_{y2}}$ can be analytically obtained by Eq. (13) when ${\eta _1},{u_1},{k_{11}},{k_{12}},{\beta _1},{\phi _1}$ are given. Meanwhile, ${\eta _2},{u_2},{k_{21}},{k_{22}},{\beta _2},{\phi _2}$ and $\Delta {a_2},\Delta {i_2},\Delta {\Omega _2},\Delta {e_{x2}},\Delta {e_{y2}}$ correspond by Eq. (12). Thus, if we can obtain the inverse solution of Eq. (12), the constraint of Eq. (13) can be eliminated from the optimization model and only $\Delta {t_1},\Delta {t_2}$,${\eta _1},{u_1},{k_{11}},{k_{12}},{\beta _1}$ and ${\phi _1}$ will be retained as unknown parameters. \\
Solving ${\eta _2},{u_2},{k_{21}},{k_{22}},{\beta _2},{\phi _2}$ by $\Delta {a_2},\Delta {i_2},\Delta {\Omega _2},\Delta {e_{x2}},\Delta {e_{y2}}$ is a boundary value problem like the Lambert's problem. First, ${u_2}$ can be directly obtained by Eq. (12):
\begin{equation}
{u_2} = \arctan (\Delta {\Omega _2}\sin {i_0},\Delta {i_2})
\end{equation}
Then, defining $C$ and $D$ as temporary variables, we get:
\begin{equation}
\left\{\begin{array}{l}
C = \frac{{({k_{21}}k_{21}^T - {\eta _2}{k_{22}}k_{22}^T)\alpha \cos {\beta _2}\Delta {t_2}}}{{{V_0}}} = \Delta {e_{x2}}\cos {u_2} + \Delta {e_{y2}}\sin {u_2}\\
D = \frac{{({k_{21}}k_{21}^T + {k_{22}}k_{22}^T)\alpha \sin {\beta _2}\sin {\phi _2}\Delta {t_2}}}{{2{V_0}}} = \Delta {e_{x2}}\sin {u_2} - \Delta {e_{y2}}\cos {u_2}
\end{array}\right.
\end{equation}
Then, ${\phi _2}$ can be obtained:
\begin{equation}
{\phi _2} = \arctan (D,\sqrt {\Delta i_2^2 + \Delta \Omega _2^2{{\sin }^2}{i_0}} )
\end{equation}
To solve ${\eta _2},{\beta _2},{k_{21}},{k_{22}}$, two conditions are considered, and the best solution will be reserved as the optimal solution:\\
(1) ${\eta _2} =  - 1$\\
In this condition, ${\beta _2}$ can be directly obtained by Eq. (21):
\begin{equation}
{\beta _2} = \arctan (\frac{{2D}}{{\sin {\phi _2}}},C)
\end{equation}
Thus, the equations of ${k_{21}},{k_{22}}$ in Eq.(21) are
\begin{equation}
\left\{\begin{array}{l}
{k_{21}}k_{21}^T + {k_{22}}k_{22}^T = \frac{2}{\pi }(\sin \frac{\pi }{2}{k_{21}} + \sin \frac{\pi }{2}{k_{22}}) = \frac{{C{V_0}}}{{\Delta t_2^{}\alpha \cos {\beta _2}}} = E\\
{k_{21}} - {k_{22}} = \frac{{\Delta {a_2}{V_0}}}{{{a_0}\Delta t_2^{}\alpha \cos {\beta _2}}} = F
\end{array}\right.
\end{equation}
where E and F are temporary variables and Eq. (24) can be rewritten as
\begin{equation}
\begin{array}{l}
\sin \frac{\pi }{2}{k_{21}} + \sin \frac{\pi }{2}({k_{21}} - F) = \frac{\pi }{2}E\\
 \Rightarrow (1 + \cos \frac{\pi }{2}F)\sin \frac{\pi }{2}{k_{21}} - \sin \frac{\pi }{2}F\cos \frac{\pi }{2}{k_{21}} = \frac{\pi }{2}E\\
 \Rightarrow \sin (\frac{\pi }{2}{k_{21}} - G) = \frac{{\frac{\pi }{2}E}}{{\sqrt {2 + 2\cos \frac{\pi }{2}F} }}
\end{array}
\end{equation}
where $G = \arctan (\sin \frac{\pi }{2}F,1 + \cos \frac{\pi }{2}F)$ is a temporary variable. Hence, when $\left| {\frac{{\frac{\pi }{2}E}}{{\sqrt {2 + 2\cos \frac{\pi }{2}F} }}} \right| > 1$, there is no solution of ${k_{21}}$ and ${k_{22}}$. Otherwise, there are two solutions:
\begin{equation}
\begin{array}{l}
\frac{\pi }{2}{k_{21}} = \arcsin (\frac{{\frac{\pi }{2}E}}{{\sqrt {2 + 2\cos \frac{\pi }{2}F} }}) + G\\
\frac{\pi }{2}{k_{21}} = \pi  - \arcsin (\frac{{\frac{\pi }{2}E}}{{\sqrt {2 + 2\cos \frac{\pi }{2}F} }}) + G
\end{array}
\end{equation}
One can calculate the objective function of two solutions and validate the constraint ${k_{12}} + {k_{22}} \leq 2$, and then choose the feasible solution that minimizes $J$.\\
(2) ${\eta _2} = 1$\\
In this condition, ${\beta _2}$, ${k_{21}}$ and ${k_{22}}$ should be solved jointly by Eq.(12) and Eq.(21):
\begin{equation}
\left\{\begin{array}{l}
({k_{21}}k_{21}^T - {k_{12}}k_{12}^T)\cos {\beta _2} = \frac{{C{V_0}}}{{\Delta {t_2}\alpha }}\\
({k_{21}}k_{21}^T + {k_{12}}k_{12}^T)\sin {\beta _2} = \frac{{2D{V_0}}}{{\Delta {t_2}\alpha \sin {\phi _2}}}\\
({k_{21}} + {k_{22}})\cos {\beta _2} = \frac{{\Delta {a_2}{V_0}}}{{{a_0}\Delta {t_2}\alpha }}
\end{array}\right.
\end{equation}
The nonlinear solving package Minpack \citep{23} is adopted to solve the equations, and the approximate initial values can be set by four cases:\\
When ${k_{21}} \leq 1$ and ${k_{22}}\leq 1$, one can assume $\sin (\frac{{\pi {k_{21}}}}{2}) \approx \frac{{\pi {k_{21}}}}{2},\sin (\frac{{\pi {k_{22}}}}{2}) \approx \frac{{\pi {k_{22}}}}{2}$. Then, Eq.(27) can be simplified and solved as:
\begin{equation}
\begin{array}{l}
\left\{ \begin{array}{l}
({k_{21}} - {k_{12}})\cos {\beta _2} = \frac{{C{V_0}}}{{\Delta {t_2}\alpha }}\\
({k_{21}} + {k_{12}})\sin {\beta _2} = \frac{{2D{V_0}}}{{\Delta {t_2}\alpha \sin {\phi _2}}}\\
({k_{21}} + {k_{22}})\cos {\beta _2} = \frac{{\Delta {a_2}{V_0}}}{{{a_0}\Delta {t_2}\alpha }}
\end{array} \right. \Rightarrow \\
\left\{ \begin{array}{l}
{\beta _2} = \arctan (\frac{{2D{V_0}}}{{\Delta {t_2}\alpha \sin {\phi _2}}},\frac{{\Delta {a_2}{V_0}}}{{{a_0}\Delta {t_2}\alpha }})\\
{k_{21}} = \frac{{\Delta {a_2}{V_0}}}{{2{a_0}\Delta {t_2}\alpha \cos {\beta _2}}} + \frac{{C{V_0}}}{{2\Delta {t_2}\alpha }}\\
{k_{22}} = \frac{{\Delta {a_2}{V_0}}}{{2{a_0}\Delta {t_2}\alpha \cos {\beta _2}}} - \frac{{C{V_0}}}{{2\Delta {t_2}\alpha }}
\end{array} \right.
\end{array}
\end{equation}
When ${k_{21}} > 1$ and ${k_{22}}\leq 1$, one can assume $\sin (\frac{{\pi {k_{21}}}}{2}) \approx \frac{{\pi (2 - {k_{21}})}}{2},\sin (\frac{{\pi {k_{22}}}}{2}) \approx \frac{{\pi {k_{22}}}}{2}$. Then, Eq.(27) can be simplified and solved as:
\begin{equation}
\begin{array}{l}
\left\{ \begin{array}{l}
(2 - {k_{21}} - {k_{12}})\cos {\beta _2} = \frac{{C{V_0}}}{{\Delta {t_2}\alpha }}\\
(2 - {k_{21}} + {k_{12}})\sin {\beta _2} = \frac{{2D{V_0}}}{{\Delta {t_2}\alpha \sin {\phi _2}}}\\
({k_{21}} + {k_{22}})\cos {\beta _2} = \frac{{\Delta {a_2}{V_0}}}{{{a_0}\Delta {t_2}\alpha }}
\end{array} \right. \Rightarrow \\
\left\{ \begin{array}{l}
{\beta _2} = \arccos (\frac{{C{V_0}}}{{2\Delta {t_2}\alpha }} + \frac{{\Delta {a_2}{V_0}}}{{2{a_0}\Delta {t_2}\alpha }})\\
{k_{21}} = (\frac{{\Delta {a_2}{V_0}}}{{{a_0}\Delta {t_2}\alpha \cos {\beta _2}}} + 2 - \frac{{2D{V_0}}}{{\Delta {t_2}\alpha \sin {\alpha _2}\sin {\beta _2}}})/2\\
{k_{22}} = (\frac{{\Delta {a_2}{V_0}}}{{{a_0}\Delta {t_2}\alpha \cos {\beta _2}}} - 2 + \frac{{2D{V_0}}}{{\Delta {t_2}\alpha \sin {\alpha _2}\sin {\beta _2}}})/2
\end{array} \right.
\end{array}
\end{equation}
When ${k_{21}} \leq 1$ and ${k_{22}}> 1$, one can assume $\sin (\frac{{\pi {k_{21}}}}{2}) \approx \frac{{\pi {k_{21}}}}{2},\sin (\frac{{\pi {k_{22}}}}{2}) \approx \frac{{\pi (2 - {k_{22}})}}{2}$. Then, Eq.(27) can be simplified and solved as:
\begin{equation}
\begin{array}{l}
\left\{ \begin{array}{l}
(2 - {k_{21}} - 2 + {k_{12}})\cos {\beta _2} = \frac{{C{V_0}}}{{\Delta {t_2}\alpha }}\\
(2 - {k_{21}} + 2 - {k_{12}})\sin {\beta _2} = \frac{{2D{V_0}}}{{\Delta {t_2}\alpha \sin {\phi _2}}}\\
({k_{21}} + {k_{22}})\cos {\beta _2} = \frac{{\Delta {a_2}{V_0}}}{{{a_0}\Delta {t_2}\alpha }}
\end{array} \right. \Rightarrow \\
\left\{ \begin{array}{l}
{\beta _2} = \arccos ( - \frac{{C{V_0}}}{{2\Delta {t_2}\alpha }} - \frac{{\Delta {a_2}{V_0}}}{{2{a_0}\Delta {t_2}\alpha }})\\
{k_{21}} = (\frac{{\Delta {a_2}{V_0}}}{{{a_0}\Delta {t_2}\alpha \cos {\beta _2}}} + \frac{{2D{V_0}}}{{\Delta {t_2}\alpha \sin {\alpha _2}\sin {\beta _2}}} - 2)/2\\
{k_{22}} = (\frac{{\Delta {a_2}{V_0}}}{{{a_0}\Delta {t_2}\alpha \cos {\beta _2}}} - \frac{{2D{V_0}}}{{\Delta {t_2}\alpha \sin {\alpha _2}\sin {\beta _2}}} + 2)/2
\end{array} \right.
\end{array}
\end{equation}
When ${k_{21}} > 1$ and ${k_{21}} > 1$, ${k_{21}} + {k_{22}} > 2$ cannot be satisfied and there is no solution for this case. \\
Eq. (28), Eq. (29), and Eq. (30) can be used as initial values to solve ${\beta _2}$, ${k_{21}}$ and ${k_{22}}$ separately. Further, the constraints on ${k_{21}}$ and ${k_{22}}$ should be validated, and the feasible solution that minimizes $J$ will be retained. \\
Above all, the solving process from $\Delta {a_2},\Delta {i_2},\Delta {\Omega _2},\Delta {e_{x2}},\Delta {e_{y2}}$ to ${\eta _2},{u_2},{k_{21}},{k_{22}},{\beta _2},{\phi _2}$ is obtained. If there is no feasible solution obtained after this process, the given $\Delta {a_2},\Delta {i_2},\Delta {\Omega _2},\Delta {e_{x2}},\Delta {e_{y2}}$ are out of the reachable domain of low-thrust, and a punish term should be added to the objective function (Eq. (18)).\\

\subsection{Optimization and parameters correction }
\begin{figure*}
\center
\begin{minipage}[t]{120mm}
\vspace {2mm}
\centerline{\includegraphics[scale=0.7]{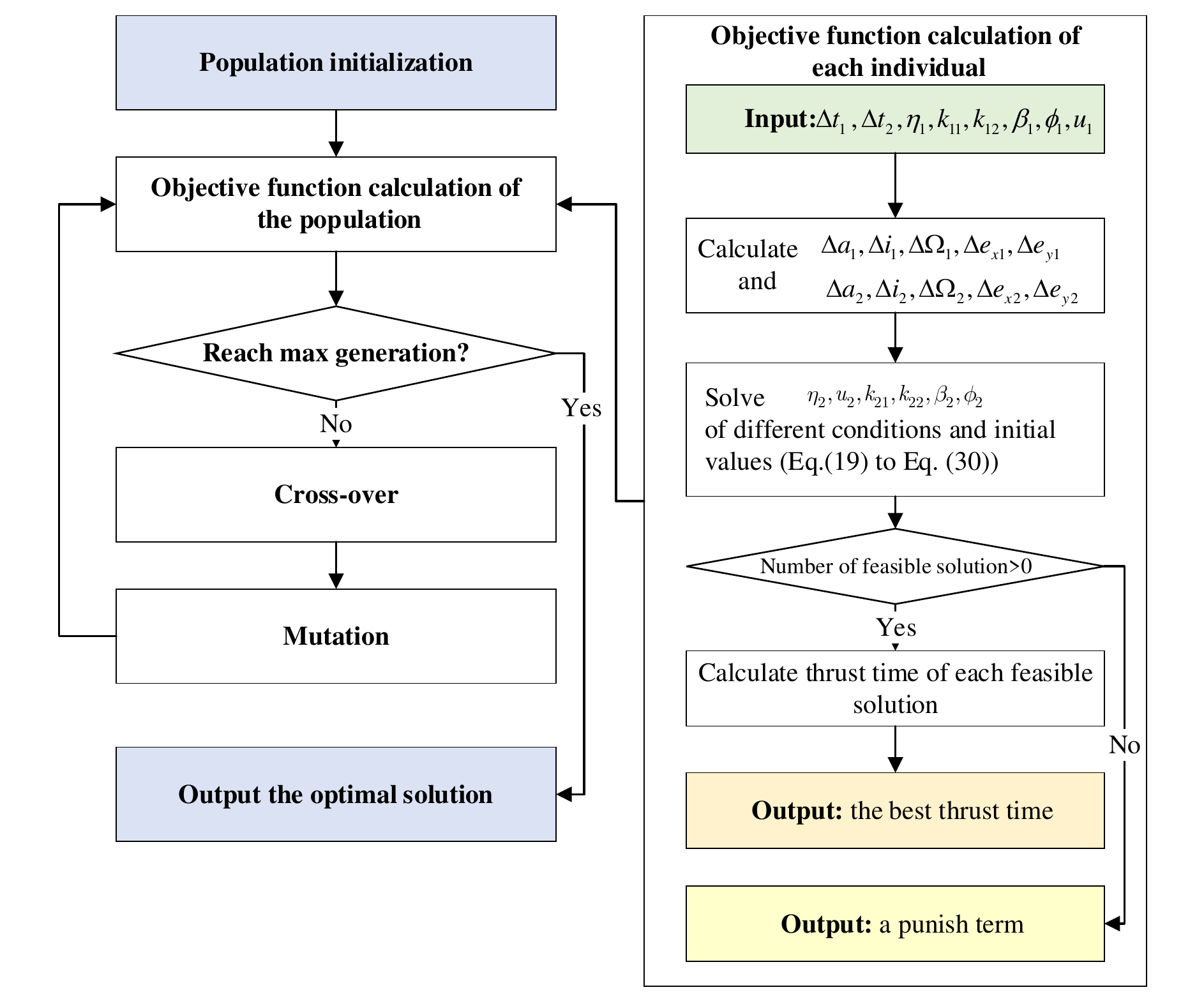}}
\vspace {-1mm}
\caption{Flowchart of the optimization process.}
\label{fig:fig5}
\vspace {-3mm}
\end{minipage}
\end{figure*}

\begin{figure}
\centering
\includegraphics[scale=0.8]{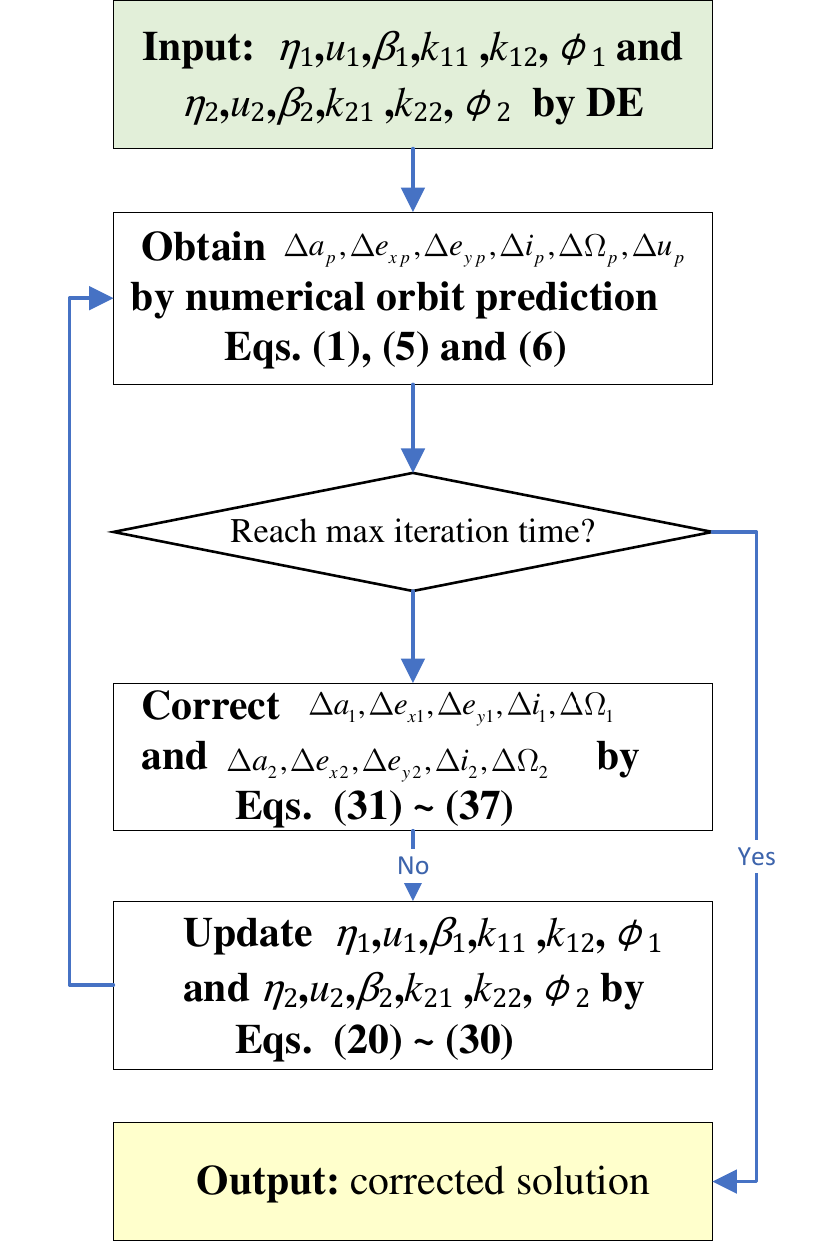}
\caption{Flowchart of the correction process.}
\label{fig:fig6}
\end{figure}

After applying the algorithm in Section III.D, there are 8 unknowns. Differential evolution (DE) algorithm \citep{24} can be used to solve this optimization problem. DE is an efficient evolutionary algorithm for the optimization of continuous variables. Details of DE are beyond the scope of the study and one can refer to \citep{24}. The process is illustrated in Fig. \ref{fig:fig5}.\\
It should be noted that due to the approximations in the optimization model if we substitute the optimal law of thrust in the numerical dynamic model (Eq. (1), Eq. (5) and Eq. (6)), there may be errors between given target orbit and the predicted orbit after control. Let $\Delta {a_p},\Delta {e_x}_p,\Delta {e_y}_p,\Delta {i_p},\Delta {\Omega _p},\Delta {u_p}$ denote the numerical errors of semimajor axis, eccentricity (two dimensions), inclination, RAAN and argument of latitude, an correction to the solution obtained by DE is proposed as follows. We first calculate the analytical corrections to the changes in orbit elements ($\Delta {a_1},\Delta {e_x}_1,\Delta {e_y}_1,\Delta {i_1},\Delta {\Omega _1}$, and $\Delta {a_2},\Delta {e_x}_2,\Delta {e_y}_2,\Delta {i_2},\Delta {\Omega _2}$) and then update the parameters of thrust (${\eta _1},{u_1},{k_{11}},{k_{12}},{\beta _1},{\phi _1}$ and ${\eta _2},{u_2},{k_{21}},{k_{22}},{\beta _2},{\phi _2}$) by Eqs. (19) to (30), respectively. \\
First, let $\Delta a_1^c$ and $\Delta a_2^c$ denote the corrections to $\Delta a_1^{}$ and $\Delta a_2^{}$. According to Eq. (13) and (17), we can obtain:
\begin{equation}
\left\{\begin{array}{l}
\Delta a_1^c + \Delta a_2^c = \Delta a_{1p}^{}\\
 - \frac{{3{n_0}}}{{2{a_0}}}\Delta a_1^c(\Delta t - \frac{{\Delta {t_1}}}{2}) - \frac{{3{n_0}}}{{2{a_0}}}\Delta a_2^c(\Delta {t_2} - \frac{{\Delta {t_2}}}{2}) = \Delta {u_p}
\end{array}\right.
\end{equation}
where $ - \frac{{3{n_0}}}{{2{a_0}}}\Delta a_1^c$ and $ - \frac{{3{n_0}}}{{2{a_0}}}\Delta a_2^c$ are changes in angular velocity by $\Delta a_1^c$ and $\Delta a_2^c$. Then, $\Delta a_1^c$ and $\Delta a_2^c$ can be solved by Eq.(31):
\begin{equation}
\left\{\begin{array}{l}
\Delta a_1^c = \frac{{\frac{{\Delta {t_2}}}{2}\Delta a_{1p}^{} + \frac{{2{a_0}\Delta {u_p}}}{{3{n_0}}}}}{{(\Delta {t_2} - \Delta t)}}\\
\Delta a_2^c = \Delta a_{1p}^{} - \Delta a_1^c
\end{array}\right.
\end{equation}
Second, let $\Delta i_1^c$ and $\Delta i_2^c$ denote the correction to $\Delta i_1^{}$ and $\Delta i_2^{}$. $u_1^{}$ and $u_2^{}$ are assumed unchanged after the correction, which is more convenient to derive analytical euqations of $\Delta i_1^c$ and $\Delta i_2^c$ \citep{21}. According to Eq. (13) and (17), we can obtain:
\begin{equation}
\left\{\begin{array}{l}
\Delta i_1^c + \Delta i_2^c = \Delta i_p^{}\\
\Delta \Omega _1^c + \Delta \Omega _2^c + {{\dot \Omega }_0}( - 3.5\frac{{\Delta a_1^c}}{{{a_0}}} - \tan {i_0}\Delta i_1^c)(\Delta t - \frac{{\Delta {t_1}}}{2})\\ + {{\dot \Omega }_0}( - 3.5\frac{{\Delta a_2^c}}{{{a_0}}} - \tan {i_0}\Delta i_2^c)(\frac{{\Delta {t_2}}}{2}) = \Delta \Omega _p^{}
\end{array}\right.
\end{equation}
where ${\dot \Omega _0}( - 3.5\frac{{\Delta a_1^c}}{{{a_0}}} - \tan {i_0}\Delta i_1^c)$ and ${\dot \Omega _0}( - 3.5\frac{{\Delta a_2^c}}{{{a_0}}} - \tan {i_0}\Delta i_2^c)$ means the changes in RAAN caused by the corrections to semimajor axis and inclination. Because $\Delta \Omega _1^c$ and $\Delta \Omega _2^c$ can be expressed by $\frac{{\tan {u_1}}}{{\sin {i_0}}}\Delta i_1^c$ and $\frac{{\tan {u_2}}}{{\sin {i_0}}}\Delta i_2^c$, Eq.(33) can be transformed to binary linear equations of $\Delta i_1^c$ and $\Delta i_2^c$, and can be solved as:
\begin{equation}
\left\{ {\begin{array}{*{20}{l}}
{\Delta i_1^c + \Delta i_2^c = \Delta i_p^{}}\\
{\frac{{\tan {u_1}}}{{\sin {i_0}}}\Delta i_1^c + \frac{{\tan {u_2}}}{{\sin {i_0}}}\Delta i_2^c = S}
\end{array}} \right. \Rightarrow \left\{ {\begin{array}{*{20}{l}}
{\Delta i_1^c = \frac{{R\Delta i_p^{} - S}}{{R - Q}}}\\
{\Delta i_2^c = \Delta i_p^{} - \Delta i_1^c}
\end{array}} \right.
\end{equation}
where Q, R and S are temporary variables:
\begin{equation}
\left\{\begin{array}{l}
Q = (\frac{{\tan {u_1}}}{{\sin {i_0}}} - \tan {i_0}{{\dot \Omega }_0}(\Delta t - \frac{{\Delta {t_1}}}{2}))\\
R = (\frac{{\tan {u_2}}}{{\sin {i_0}}} - \tan {i_0}{{\dot \Omega }_0}\frac{{\Delta {t_2}}}{2})\\
S = \Delta \Omega _p^{} + {{\dot \Omega }_0}(3.5\frac{{\Delta a_1^c}}{{{a_0}}})(\Delta t - \frac{{\Delta {t_1}}}{2}) + {{\dot \Omega }_0}(3.5\frac{{\Delta a_2^c}}{{{a_0}}})(\frac{{\Delta {t_2}}}{2})
\end{array}\right.
\end{equation}
Thus, $\Delta \Omega _1^c$ and $\Delta \Omega _2^c$ are also obtained.\\
Third, divide $\Delta e_{xp}^{}$ and $\Delta {e_{yp}}$ into two parts as the corrections to $\Delta {e_{x1}},\Delta {e_{y1}},\Delta {e_{x2}}$ and $\Delta e_{y2}^{}$ according to the ratio of thrust time in the first and third stages:
\begin{equation}
\left\{\begin{array}{l}
\Delta {e_{x1}^c} = \chi \Delta {e_{xp}}\\
\Delta {e_{y1}^c} = \chi \Delta {e_{yp}}\\
\Delta {e_{x2}^c} = (1 - \chi )\Delta {e_{xp}}\\
\Delta {e_{y2}^c} = (1 - \chi )\Delta {e_{yp}}
\end{array}\right.
\end{equation}
where the superscript 'c' represents corrention term and $\chi  = \frac{{\left| {{k_{11}}} \right| + \left| {{k_{12}}} \right|}}{{\left| {{k_{11}}} \right| + \left| {{k_{12}}} \right| + \left| {{k_{21}}} \right| + \left| {{k_{22}}} \right|}}$ represents the ratio of thrust time in the first transfer stage and total thrust time. Eq. (36) means the corrections of numerical eccentricity error that approximately allocated to the first and third transfer stage are proportionate to their thrust time. Note that the correction to $\Delta {e_{x1}}$ and $\Delta {e_{y1}}$ should consider the natural drift of argument of perigee during the transfer. Therefore, according to Fig. \ref{fig:fig3}, $\Delta {e_{x1}}$ and $\Delta {e_{y1}}$ should be recalculated to correct the drift angle $\Delta {\omega _d}$ by Eq. (15):\\
\begin{equation}
\left\{\begin{array}{l}
{u_1}^c = \arctan (\Delta {e_{y1}}^c,\Delta {e_{x1}}^c)\\
\Delta {e_1}^c = \sqrt {{{(\Delta {e_{x1}}^c)}^2} + {{(\Delta {e_{y1}}^c)}^2}} \\
\Delta {e_{x1}}^c = \Delta {e_1}^c\cos ({u_1}^c - \Delta {\omega _d})\\
\Delta {e_{y1}}^c = \Delta {e_1}^c\cos ({u_1}^c - \Delta {\omega _d})
\end{array}\right.
\end{equation}
where ${u_1}^c$ is the argument of latitude of the eccentricity correction and $\Delta {e_1}^c$ is the magnitude. \\
Above all, $\Delta {a_1},\Delta {i_1},\Delta {\Omega _1},\Delta {e_{x1}},\Delta {e_{y1}}$ and $\Delta {a_2},\Delta {i_2},\Delta {\Omega _2},\Delta {e_{x2}},\Delta {e_{y2}}$ have been correct and the parameters ${\eta _1},{u_1},{\beta _1},{\phi _1},{k_{11}},{k_{12}}$ and ${\eta _2},{u_2},{\beta _2},{\phi _2},{k_{21}},{k_{22}}$ can be updated by solving Eqs. (20) ~ (30), respectively. The correction process can be repeated multiple times to obtain high-precision trajectory. The flow chart is illustrated in Fig.\ref{fig:fig6}. \\

\section{Simulation Result}
In this study, a transfer between two space debris is analyzed. The initial and target orbits of the spacecraft are listed in Table \ref{table:tab1}. The transfer duration is 20 days and the constant acceleration is $6\times10^{-4} m/s^2$. The population of DE is set to 50, and the max generation is 800. Meanwhile, the crossover operator is 'Ran2Bestexp' \citep{24} with a probability of 0.8 and the mutation probability is set to 0.8. \\
\begin{table*}[hbt!]
\centering
\caption{Detail of orbits}
\label{table:tab1}
\begin{tabular}{|l|l|l|l|l|l|l|l|}
\hline
{ }& Orbit	& $a (m)$	& $e$	& $i$ (deg)	& $\Omega$ (deg)	& $\omega$ (deg)	& $M$ (deg)\\
\hline
\multirow{2}*{Initial}& 	Mean	& 7157398	& 0.01521	& 98.6435	& 152.508	& 20.285	& 341.629\\
\cline{2-8}
~&	Osculating	& 7166678	& 0.01566	& 98.637	& 152.507	& 19.818	& 342.100\\
\hline
\multirow{2}*{Target}	& Mean 	& 7111954	& 0.00721	& 97.4512	& 151.175	& 44.985	& 59.376\\
\cline{2-8}
~&	Osculating	& 7103971	& 0.00678	& 97.455	& 151.178	& 32.638	& 71.695\\
\hline
\end{tabular}
\end{table*}
\subsection{Optimal solution}
\begin{figure*}
\center
\begin{minipage}[t]{120mm}
\vspace {2mm}
\centerline{\includegraphics[scale=0.7]{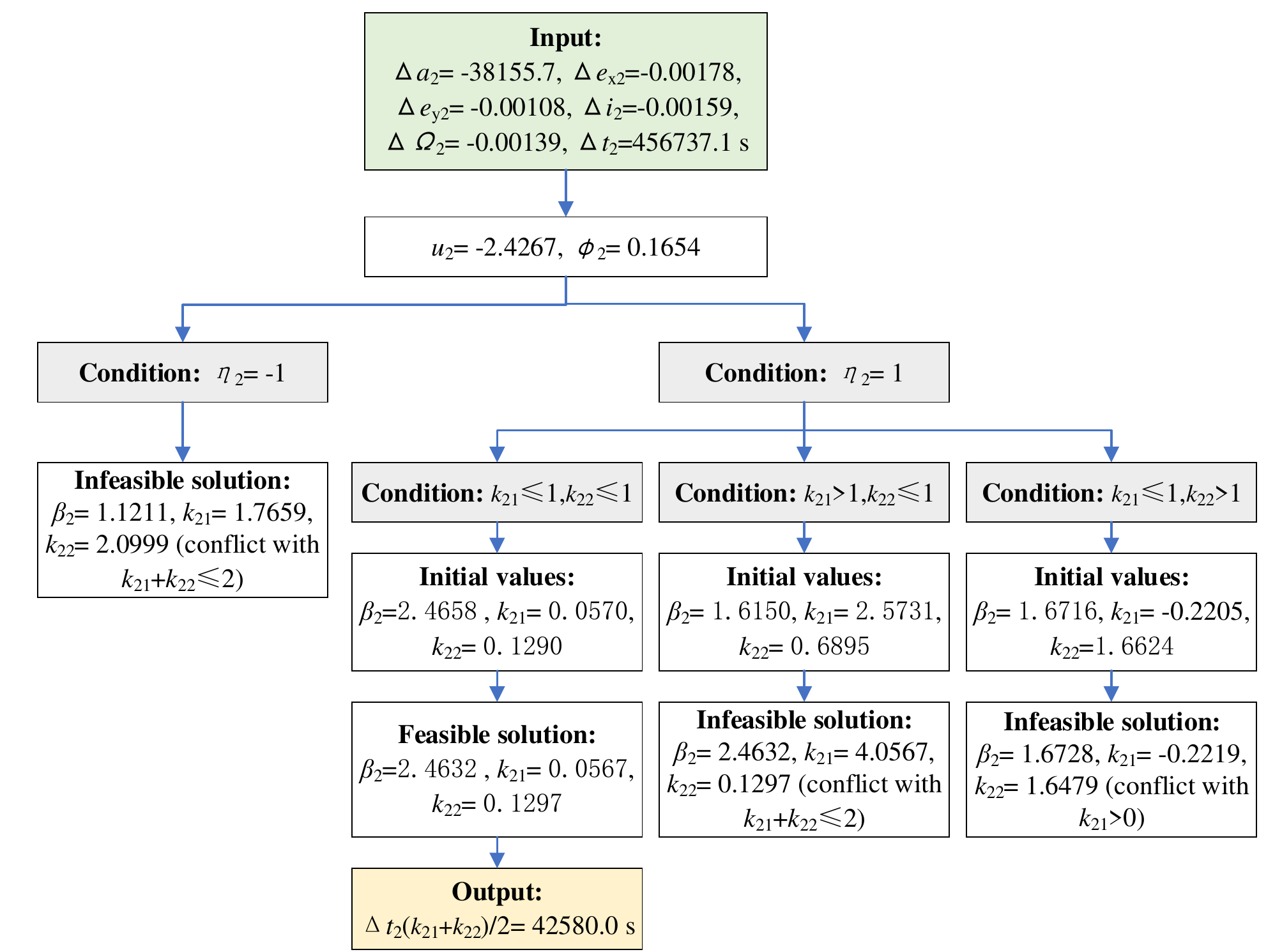}}
\vspace {-1mm}
\caption{Process of the sub boundary value problem.}
\label{fig:fig7}
\vspace {-3mm}
\end{minipage}
\end{figure*}

\begin{figure*}
\centering
\begin{minipage}[t]{120mm}
\vspace {2mm}
\centerline{\includegraphics[scale=0.7]{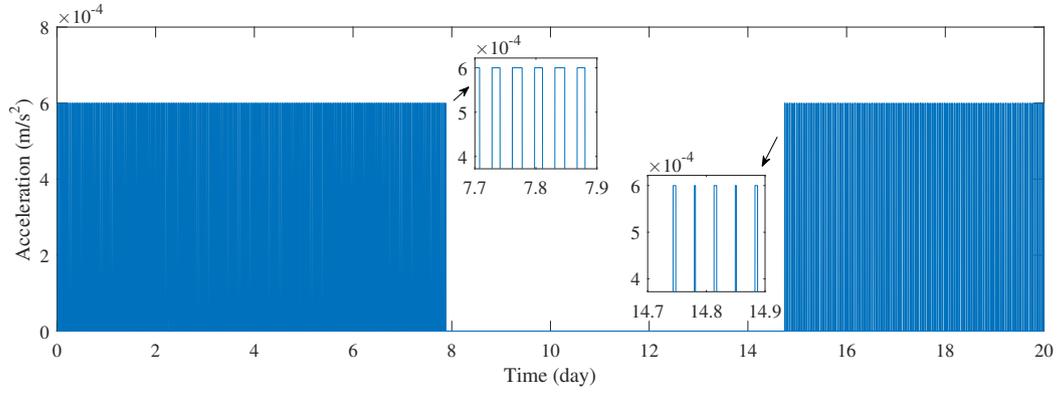}}
\vspace {-1mm}
\caption{Fuel-optimal law of thrust.}
\label{fig:fig8}
\vspace {-3mm}
\end{minipage}
\end{figure*}

\begin{figure*}
\centering
\begin{minipage}[t]{120mm}
\vspace {2mm}
\centerline{\includegraphics[scale=0.7]{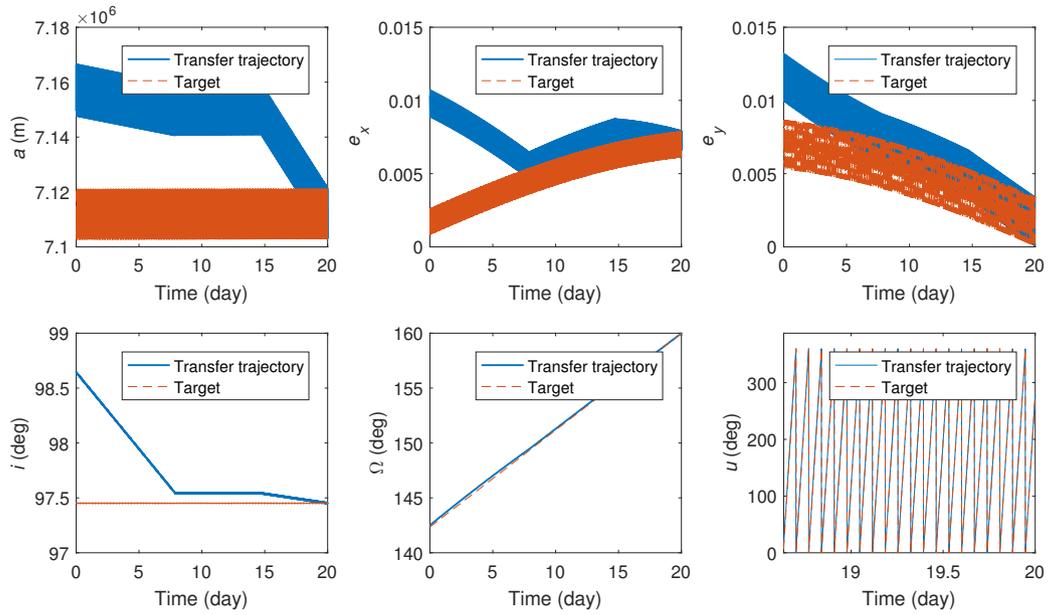}}
\vspace {-1mm}
\caption{History of orbit elements during the transfer.}
\label{fig:fig9}
\vspace {-3mm}
\end{minipage}
\end{figure*}

The optimal $\Delta {t_1},\Delta {t_2},{\eta _1},{u_1},{\beta _1},{\alpha _1},{k_{11}},{k_{12}}$,${\eta _2},{u_2},{\beta _2},{\phi _2},{k_{21}}$, and ${k_{22}}$ obtained by DE are [7.8908, 5.2863, -1, -2.7460, 1.3779, 0.0141, 0.3691, 0.4660, 1, -2.4267, 2.4632, 0.1654, 0.0567, 0.1297]. The total length of thrust-on arcs is 3.78 days, and the equivalent velocity increment is 196.35 m/s. The calculation takes less than 2 s on a personal computer (CPU: AMD Ryzen7 4.2 GHz). \\
The process of the sub boundary value problem in Section III.D corresponding to the optimal solution is illustrated in Fig.\ref{fig:fig7}. In summary, the algorithm can correctly check the low-thrust reachability and quickly obtain ${\eta _2},{u_2},{\beta _2},{\phi _2},{k_{21}},{k_{22}}$ by given $\Delta {a_2},\Delta {i_2},\Delta {\Omega _2},\Delta {e_{x2}},\Delta {e_{y2}}$. Repeating calculations proved that less than 20 shootings are required for Minpack to solve Eq. (27) with initial values by Eqs. (28-30).\\
\begin{table*}[hbt!]
\centering
\caption{Numerical errors of orbit elements}\label{table:tab2}
\begin{tabular}{|l|l|l|l|l|l|l|}
\hline
Step    	&$\Delta a_p (m)$	&$\Delta e_{xp}$	&$\Delta e_{yp}$	&$\Delta i_p (deg)$	 &$\Delta\Omega_p (deg)$	&$\Delta u_p (deg)$\\
\hline
0	&489.1315	&-0.0035	&-0.0089	&-12.2485	&0.00105	&0.00163\\
\hline
1	&-94.4675	&0.00043	&0.01118	&1.5021	&0.000261	&-0.00011\\
\hline
2	&-18.1261	&-9.47E-05	&-0.00149	&0.2739	&-8.58E-06	&-3.85E-05\\
\hline
3	&1.76396	&-1.21E-05	&4.21E-06	&-0.0371	&-5.76E-06	&5.23E-07\\
\hline
4	&0.45304	&2.02E-06	&3.69E-05	&-0.0093	&-7.10E-08	&7.79E-07\\
\hline
5	&0.01925	&7.73E-07	&1.81E-06	&-2.36E-04	&1.09E-07	&1.73E-08\\
\hline
\end{tabular}
\end{table*}
The errors after each correction step are also detailed in Table \ref{table:tab2}. After five steps, the errors are close to zero and can be ignored (the position error is less than 10 m and the velocity error is less than 0.03 m/s). Finally the optimal 14 parameters are [7.8908, 5.2863, -1, -2.7473, 1.3809, 0.0906, 0.3748, 0.4650, 1, -2.4307, 2.4632, 0.1846, 0.0636, 0.1239]. The optimal thrust is illustrated in Fig. \ref{fig:fig8}. Differing from the symmetrical thrust strategy in \citep{21}, the lengths of two thrust-on arcs in each revolution are not the same to make use of the tangential acceleration to change eccentricity. The history of the orbit elements is illustrated in Fig. \ref{fig:fig9}, which indicates that the optimal law of thrust firstly should partly decrease the semimajor axis and inclination to achieve the optimal drift rates of RAAN and the argument of perigee. Finally, after a long duration of natural drift, the thrust is employed again to control the spacecraft to rendezvous with the target orbit. The equivalent velocity increment (197.48 m/s) is very close to the minimum velocity increment (196.10 m/s) obtained by the indirect method \citep{6, 7, 8} after multiple calculations (repeating with different randomly generated co-states to obtain the global optimal solution), which proved the optimality of the proposed method.\\

\subsection{Analysis of different thrust accelerations}
\begin{figure*}
\centering
\includegraphics[scale=0.8]{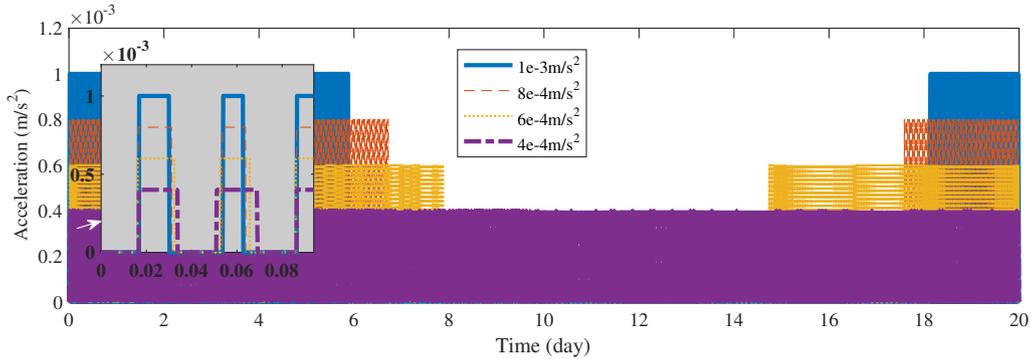}
\caption{Law of thrust of different accelerations.}
\label{fig:fig10}
\end{figure*}

\begin{figure}
\centering
\includegraphics[scale=0.7]{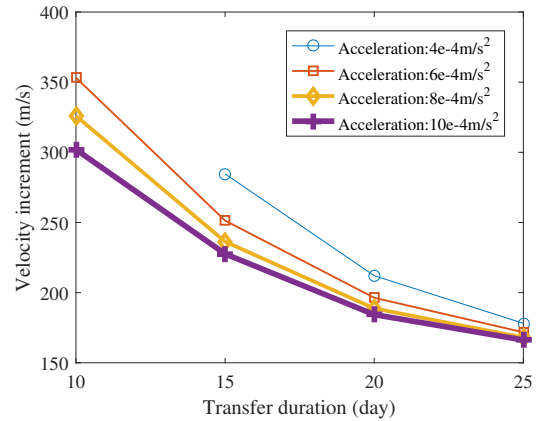}
\caption{Velocity increment of different transfer duration and acceleration.}
\label{fig:fig11}
\end{figure}

To validate the applicability of the proposed method for different thrust levels, four cases of thrust acceleration ($1\times10^{-3} m/s^2$, $8\times10^{-4} m/s^2$, $6\times10^{-4} m/s^2$, $4\times10^{-4} m/s^2$) are tested and the law of thrust are illustrated in Fig.\ref{fig:fig10}. When the thrust acceleration is larger, the optimal solution prefers longer natural drift duration to save fuel because the spacecraft can be transferred to the drift orbit faster. When the thrust acceleration is smaller, it’s more difficult to transfer to the drift orbit and thus larger velocity increment is required for direct non-coplanar control. However, it can be seen that coast arcs in the transfers arriving and departing the drift orbit are usually necessary to save propellant. It’s also obtained that the orbit transfer problem in Table \ref{table:tab1} will be infeasible when the acceleration is smaller than $2.7\times10^{-4} m/s^2$. \\
The equivalent velocity increments of different transfer durations and different thrust accelerations are illustrated in Fig.\ref{fig:fig11}, which indicates the natural drift of RAAN and argument of perigee can greatly decrease the propellant when the transfer duration is long enough. The method in this paper can well adept with trajectory optimization of different conditions. Optimization by DE could ensure the thrust parameters achieve the balance between direct control of orbit elements and indirect control via natural drift.\\

\subsection{Comparation with previous methods and discussion}
\begin{figure}
\centering
\includegraphics[scale=0.7]{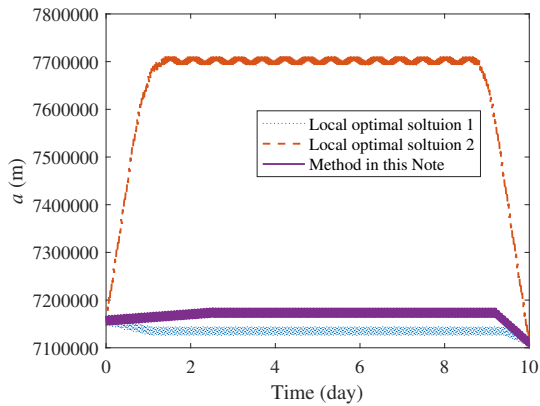}
\caption{Velocity increment of different transfer duration and acceleration.}
\label{fig:fig12}
\end{figure}

In this study, the thrust strategy is just approximately optimal because the direction of thrust acceleration is fixed and the thrust keeps the same in different revolutions. However, the solutions obtained from repeated calculations are consistent and are proved to be always very close to the best results of the indirect methods with numerical dynamics \citep{6, 7, 8} after a lot of simulations with different orbits and transfer durations. By contrast, the methods in \citep{6, 7, 8} may easily converge to locally optimal solutions when solving such long-duration perturbed orbit rendezvous problems because the number of revolutions is large and the effect of perturbations cannot be globally considered. Fig. \ref{fig:fig12} shows examples of several local optimal solutions ($\Delta t$ = 10 days) obtained by an indirect method corresponding to very large velocity increments. \\
The approximate methods in \citep{17, 18, 19, 21} are also tested using the same orbits in Table \ref{table:tab1} with different transfer durations and different thrust accelerations. Taking the best results of the indirect method as a benchmark, the comparisons of precision and efficiency between different methods (using the same computer) are detailed in Table \ref{table:tab3}, which demonstrates the performance of the proposed method. For the case in Table \ref{table:tab1}, the relative errors of methods in \citep{18, 19, 21} are greater than 30\% because the eccentricity difference is about 0.01, which requires additional velocity increment for orbit rendezvous. The calculation of the proposed method is much less than existing indirect methods requiring hundreds of shooting processes and the thrust law is also much simpler for engineering practice. Besides, the results are also closer to the global optimal solution than existing approximate methods. \\
\begin{table*}[hbt!]
\centering
\caption{Comparison of different methods}\label{table:tab3}
\resizebox{\linewidth}{!}{
\begin{tabular}{|c|c|c|}
\hline
Method	&Calculation time (s) &Maximum relative error\\
\hline
This study	&2	&3.5\%\\
\hline
Numerical method in \citep{6,7,8}	&\textgreater 14400	&Used as benchmark\\
\hline
Approximate iteration method in \citep{17}	&0.01	&7.9\%\\
\hline
Approximate methods \citep{18,19,21}	&1	&Only for circular orbits\\
\hline
\end{tabular}
}
\end{table*}

\section{Conclusion}
A priori fuel-optimal thrust strategy is proposed to simplify the trajectory optimization of orbit rendezvous with low eccentricities into a parametric optimization problem, which significantly reduces the solving complexity. By parameterizing the switch strategy and direction of the thrust, the analytical expression of the rendezvous constraint and objective function are obtained and thus a sub boundary value problem is introduced to further reduce the number of unknowns. Finally, a differential evolution algorithm is adopted to solve the simplified optimization model and an analytical correction process is proposed to eliminate the numerical errors. Simulation results and comparisons with previous methods proved this new method's efficiency and high precision for low-eccentricity orbits. The method can be well applied to premilitary analysis and high-precision trajectory optimization of missions such as in-orbit service and active debris removal in low Earth orbits.\\

\section{Acknowledgment}
This work was supported by the National Natural Science Foundation of China (No. 12002394 and No. 12172382).

\bibliographystyle{jasr-model5-names}
\biboptions{authoryear}
\bibliography{hay}

\end{document}